\documentclass[pra,twocolumn,nofootinbib,showpacs,amsmath,amssymb]{revtex4}

\usepackage{graphicx}

\newcommand\Id {\openone}
\usepackage[usenames,dvipsnames]{color}

\definecolor{LinkColor}{rgb}{0,0,.5}
\usepackage[colorlinks=true,linkcolor=BrickRed,citecolor=LinkColor,urlcolor=LinkColor]{hyperref}
\renewcommand{\emph}{\textit}

\newcommand{\half}{\frac{1}{2}}

\DeclareMathSymbol{\theta}{\mathalpha}{letters}{"23}
\DeclareMathSymbol{\phi}{\mathalpha}{letters}{"27}
\newcommand{\cc}{{\cal C}}
\newcommand{\zl}[1]{\label{eqn:#1}}
\newcommand{\zr}[1]{Eq.~(\ref{eqn:#1})}
\newcommand{\zfl}[1]{\label{fig:#1}}
\newcommand{\zfr}[1]{Fig.~\ref{fig:#1}}
\newcommand{\ztl}[1]{\label{table:#1}}
\newcommand{\ztr}[1]{Table~\ref{table:#1}}
\newcommand{\zsl}[1]{\label{sec:#1}}
\newcommand{\zsr}[1]{Sec.~\ref{sec:#1}}
\newcommand{\zal}[1]{\label{app:#1}}
\newcommand{\zar}[1]{Appendix~\ref{app:#1}}

\newcommand{\ket}[1]{\left\vert{#1}\right\rangle}
\newcommand{\bra}[1]{\left\langle{#1}\right\vert}
\newcommand{\braket}[2]{\langle{#1}\vert{#2}\rangle}

\begin{document}
\title{Mixed-state quantum transport in correlated spin networks}
\author{Ashok Ajoy}
\email{ashokaj@mit.edu}
\affiliation{Department of Nuclear Science and Engineering,
Massachusetts Institute of Technology, Cambridge, MA, USA}
\author{Paola Cappellaro}
\email{pcappell@mit.edu}
\affiliation{Department of Nuclear Science and Engineering,
Massachusetts Institute of Technology, Cambridge, MA, USA}
\begin{abstract}
Quantum spin networks can be used to transport  information  between separated
registers in a quantum information processor. To find a practical
implementation, the strict requirements of ideal models for perfect state
transfer need to be relaxed, allowing for  complex coupling topologies and
general initial states.
Here we analyze  transport in complex  quantum spin
networks in the maximally mixed state  and derive explicit conditions that should be
satisfied by propagators for perfect  state  transport. 
Using a description of the transport process as a quantum walk over the network, we show that it is necessary to phase correlate
the transport processes occurring along all the possible paths in the network.
We provide a Hamiltonian that achieves this correlation, and use it in a constructive method 
to derive engineered couplings for perfect transport in complicated network
topologies. 
\end{abstract}
\maketitle

\maketitle
\section{Introduction}
\zsl{I}
In the quest toward a scalable quantum computer~\cite{Ladd10}, a promising model comprises distributed computing units connected by passive wires that transmit quantum information~\cite{Cirac99,Oi06,Jiang07,Kimble08,Meter08}.
This architecture would provide several advantages, since the wires  require no or limited control,  easing the fabrication requirements and improving their isolation from the environment. 
For a simpler integration in a solid-state architecture, the wires can be composed of spins.
Following seminal work by Bose~\cite{Bose03}, which showed that spin chains enable transporting quantum states between the ends of the chain, 
the dynamics of quantum state transfer has been widely studied (see
Ref.~\cite{Kay10} for a review) and protocols for improving the fidelity  by
coupling
engineering~\cite{Christandl04,Albanese04,Nikolopoulos04,Nikolopoulos04b,
Gualdi08,Wojcik05, Li05,Wang11}, dual-rail topologies~\cite{Burgarth05}, active
control on the chain spins~\cite{Alvarez10} or on the end spins
only~\cite{Fitzsimons06,Burgarth07a,Caneva09,Burgarth10} have been proposed. 
Recently these studies have been extended to  mixed state spin chains~\cite{Cappellaro07l,DiFranco08c,Cappellaro11,Yao11,Kaur11},
which are more easily obtained in high-temperature laboratory settings -- making them
important protagonists in practical quantum computing.
A further challenge to experimental implementation of quantum transport is the lack of chains with the desired coupling strengths, since coupling engineering is limited by fabrication constraints and by the presence of long-range interactions.
These challenges highlight the need for a systematic study of  mixed state transport in quantum systems  beyond  chains,  including more complex network topologies. 
These topologies reflect more closely actual experimental conditions 
 as well as  systems occurring in nature. For example,
there is remarkable recent evidence~\cite{Panitchayangkoon11,Engel07} that
coherent quantum transport may be the underlying reason  for the
high efficiency (of over 99\%) of photosynthetic energy transfer~\cite{Mohseni08}.

To derive explicit conditions for perfect transport we quantify geometric
constraints on the unitary propagator that drives transport in an arbitrary
network. As a consequence, we find that transferring some mixed states in a network
in general requires fewer conditions than pure state transfer. 
Transport conditions for pure
states have been previously quantified~\cite{Kostak07}, however, our method -- relying on decomposing the propagator in
orthogonal spaces -- is fundamentally different and more suitable for mixed states.

Perfect transport occurs when the bulk of the network acts like a lens to \textit{focus} transport to its ends. 
To make  this physical picture more concrete, we describe  mixed state transport
as a continuous quantum walk over the
network~\cite{Aharonov93,Baum85,Munowitz87} that progressively populates its
nodes.
Through this formalism, we derive constructive conditions on the Hamiltonian
that results in the \textit{correlation} of transport processes  through
different possible {paths} in the network. 
The correlation of transport processes leads to their constructive interference at the position of the two end-spins, giving perfect transport. 
While similar walk models have been applied to coherent transfer before (see \cite{Mulken11} for a review), our work provides their first extension to transport involving mixed states. 

The insight gained by describing quantum transport as correlated quantum walks can be used  to construct larger
networks where perfect transport is possible. Here we show a strategy to achieve this goal by engineering the coupling strengths between different nodes
of the network to construct \textit{weighted} spin networks that support perfect mixed state transport.
Feder \cite{Feder06} had considered a similar problem for pure
states by mapping the quantum walk of $N$ spinor bosons to a single particle;
this has been extended in more recent work \cite{Krovi07,Bernasconi08,Pemberton-Ross11}.
Here we find far more relaxed weighting requirements for
mixed state transport, thanks in part to a fermionic instead of bosonic mapping. In turn, this could ease the fabrication requirements for coupling engineering.

The paper is organized as follows. In \zsr{II} we define the
problem of mixed state transport. \zsr{III} provides the geometric
conditions on the propagator
for perfect transport in arbitrary networks.  We finally present  in \zsr{IV} the quantum walk formalism, which allows the correlation of
transport processes over different paths and the construction of families
of weighted networks that support perfect transfer.

\section{Transport in mixed-state networks}
\zsl{II}
Consider an $N$-spin network ${\cal N}$, whose vertices (nodes) ${\cal V}$
represent spins and whose edges ${\cal E}=\{\alpha_{ij}\}$ describe the  couplings between  spins
$i$ and $j$ (see  \zfr{1}). The system dynamics is governed by the
Hamiltonian $H=\sum_{i<j}\alpha_{ij}H_{ij}$, where $H_{ij}$ is the operator form of the interaction.
In the most general case, a spin of ${\cal N}$ may be coupled to several others, for instance, in a dipolar coupled network, $\alpha_{ij}\sim
1/r_{ij}^3$ is a function of the distance between the spins in the network. 

We assume that we can identify two nodes, labeled 1 and $N$,  that we can
(partially) control and read out, independently from the ``\textit{bulk}'' of the network,
and thus act as  the ``\textit{end}'' spins between which transport will occur.  The rest
of the spins in the network can at most be manipulated by collective control.
This also imposes restrictions on the network initialization~\cite{DiFranco08c,Cappellaro11,Yao11}. 
To relax the requirements for the network preparation, we assume to work in the
infinite-temperature limit~\cite{Cappellaro11} -- a physical setting 
easily achievable for many experimental systems -- where the bulk spins are in   the {maximally} mixed state, $\rho \propto \Id $.
We will then consider the transport  of a slight excess polarization from node $1$ to node $N$. 
The initial  state is  $\rho_i \sim (\Id +\delta Z_1)$, where $Z_1$ is the  Pauli matrix acting on spin 1 and $\delta\ll1$ denotes the  polarization excess. Since only the traceless part of the density matrix evolves in time, we will monitor the transport from  $\rho^{\Delta}_i = Z_1$ to a desired final state  $\rho^{\Delta}_f=Z_N$. 
The \textit{fidelity} of the transport process is then defined as $F(t)={\rm
Tr}(\rho^{\Delta}(t)Z_N)/{\rm
Tr}(Z^{\dag}_1Z_1)$, with
$\rho^{\Delta}(t)=U(t)\rho^{\Delta}_iU^\dag (t)$ being the evolved state.

The polarization behaves like a wave-packet traveling over the
network ${\cal N}$ \cite{Osborne04,Yung06}. 
In most cases, the Hamiltonian $H$ drives
a rapidly dispersive evolution, where the wave-packet quickly spreads out into 
many-body correlations
among the nodes of ${\cal N}$, from which it cannot be
recovered~\cite{Munowitz87}. This is for example the case of evolution under the naturally
occurring dipolar Hamiltonian, which induces a fast-decay of the spin
polarization as measured in solid-state NMR, even if many-body correlations can
be detected at longer times~\cite{Cho05}.

In order to drive a dispersionless transport, thus ensuring perfect fidelity,
the network Hamiltonian  should satisfy very specific conditions. In this paper
we will investigate these conditions by answering the questions: 
(i) What are the possible operator forms of the Hamiltonian $H_{ij}$ for dispersionless transport? 
(ii) What are the coupling topologies  and (iii)   strengths $\alpha_{ij}$ that support perfect transport?

\begin{figure}
 \centering
\includegraphics[scale=0.46]{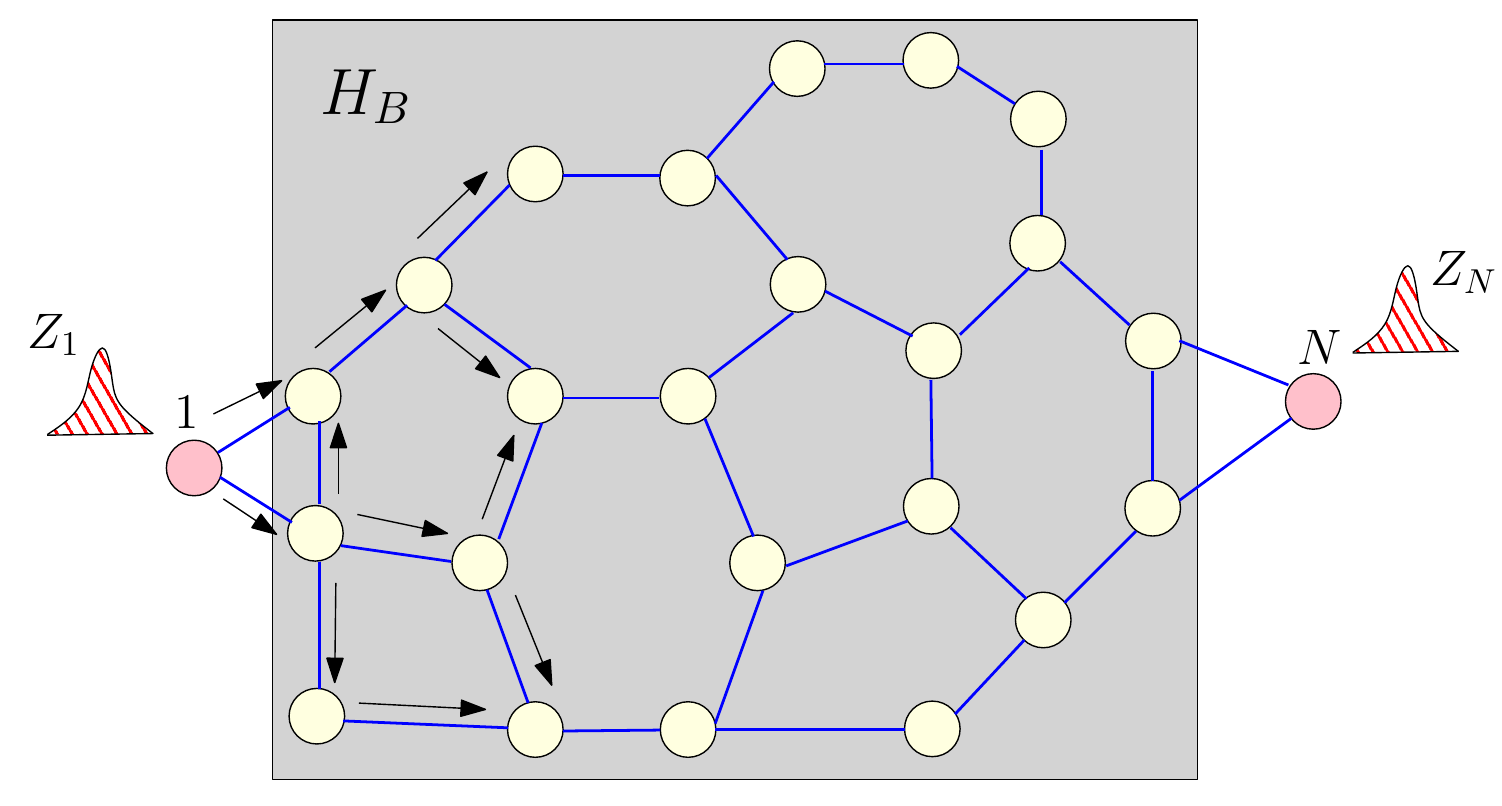}
\caption{(Color online) Transport in a spin network. The network edges
 represent  the interaction among spins (nodes).
The pink (dark) nodes represent  the {\it end} spins between which transport should occur. The shaded
region comprising  yellow (light) nodes is the {\it bulk} network, $H_B$. An
initial polarization packet (hatched) is prepared on spin 1 and allowed to propagate on
the network through various possible paths
(arrows). For perfect transport, the polarization refocuses at spin $N$.
}
\zfl{1}
\end{figure}

\section{Conditions for perfect transport}
\zsl{III}
\subsection{Fidelity of mixed-state transport}
\zsl{IIIa}
The condition for perfect transport, $F=1$, can be expressed in a  compact form by using the product-operator (PO) basis~\cite{Sorensen84}.
For the $N$-spin network system there are $2^{2N}$ basis elements, 
\begin{eqnarray}
&&\textbf{B} =\textbf{B}_1\otimes\textbf{B}_{\textrm{bulk}}\otimes\textbf{B}_N=
\{\Id ,X_{1},Y_{1},Z_{1},X_{2},\dots, X_{1}X_{2},\nonumber\\
&&\dots,Z_{1}Z_{2},X_{1}X_{2}X_{3},\dots,
\dots,Z_{1}Z_{2}\cdots Z_{N-1}Z_{N}
\},
\zl{3.1}
\end{eqnarray}
where $\textbf{B}_{1,N}$ and $\textbf{B}_{\textrm{bulk}}$ are  the basis for the end and the bulk spins, respectively.

Using the PO basis, the propagator $U(t)=e^{-iHt}$ can be represented by a vector  $\ket{U}=[c_{B_1},c_{B_2},\dots,c_{B_{4^N}}]^T$ in the
$2^{2N}$ dimensional  Hilbert-Schmidt (HS) operator space spanned by
$\textbf{B}$~\cite{Ajoy12}:
\begin{equation}
 U(t) = \textstyle\sum_i c_{B_i}(t)B_i ,\quad {\rm with}\quad
c_{B_i}=\displaystyle\frac{\textrm{Tr}(B_i^{\dagger}U)}{\textrm{Tr}(B_i^{
\dagger } B_ { i })}\nonumber
\end{equation}
From an initial state with a polarization excess on spin~1, $\rho_i^\Delta=Z_1$,
the system evolves to
\begin{equation}
 \rho^{\Delta}_f=U\rho^{\Delta}_iU^\dag=\textstyle\sum_{i,j}c_{B_i}c_{B_j}^{\ast}B_iZ_1B_j\:,
\end{equation}
 yielding the transport {fidelity} to spin $N$
\begin{eqnarray}
 F&\!=\!&\frac{1}{{\rm
Tr}(Z^{\dag}_1Z_1)}\textrm{Tr}\left[\textstyle\sum_{i,j}
c^{\ast}_{B_i}c_{B_j}B_iZ_1B_jZ_N\right]\!\!=\!\textstyle\sum_i
c_{B_i}c_{B_j}^{\ast},\nonumber\\
&& {\rm with}\ B_j= \pm Z_1Z_NB_i, \quad	 {\rm for}\quad
[B_i,Z_N]_{\mp}=0.\zl{3.2}
\end{eqnarray}
The last equation follows from the property that all elements of $\textbf{B}$,
except $\Id $, are traceless.

The fidelity derived in \zr{3.2} has a simple form
in the HS space. Note that in this operator
space, the product $B_iU$ is a linear transformation,
$\textbf{T}:\ket{U}\rightarrow\hat{P}_{B_i}\ket{U}$, where
$\hat{P}_{B_i}$ is a
permutation matrix corresponding to the action of $B_i$~\cite{Ajoy12} (here
and in the following we  denote operators in the HS space by a  hat). The
unitarity of $U$ yields  the conditions:
\begin{equation}
 \|{U}\|=\braket{U}{U}=1\qquad
\bra{U}\hat{P}_{B_i}\ket{U}=0\: ,\: B_i\neq \Id .
\zl{3.3}
\end{equation}
Let us partition the HS space in two subspaces $\cal G$, $\widetilde{\cal G}$,
spanned by the basis $G$ and $\widetilde{G}$,
\begin{equation*}
 { G} = \textbf{B}_1\otimes \textbf{B}_{\rm
bulk}\otimes\{\Id ,Z_N\},\quad 
 \widetilde{G}= \textbf{B}_1\otimes \textbf{B}_{\rm
bulk}\otimes\{X_N,Y_N\},
\end{equation*}
and note that all elements of ${G}$ commute with $Z_N$, while all elements of
$\widetilde{G}$ anti-commute with $Z_N$.
We will label by superscripts  ${{\cal G}}$ and ${\widetilde{\cal G}}$  the projections of operators in these subspaces.  
Using this partition, we can simplify the expression   for the fidelity of \zr{3.2}  to obtain
\begin{equation}
 F=\bra{U}\hat{P}_{Z_1Z_N}\hat{P}_{R}\ket{U}\:,
\zl{3.4}
\end{equation}
where $\hat{P}_{R}$ is a reflection  about ${\cal G}$ and  $\hat{P}_{Z_1Z_N}$ is block-diagonal  in the $\{{G},\widetilde{{G}}\}$
basis (since $Z_1Z_N\in {\cal G}$): 
\begin{equation}
\hat{P}_{R}=
\left[\begin{array}{cc}
\Id^{\cal G} & 0\\
0 & -\Id^{\widetilde{\cal G}}\\
\end{array}\right],\quad
\hat{P}_{Z_1Z_N}=
\left[\begin{array}{cc}
\hat{P}_{Z_1Z_N}^{\cal G} & 0\\
0 & \hat{P}_{Z_1Z_N}^{\widetilde{\cal G}}\\
\end{array}\right]
\end{equation}

Rewriting the fidelity as the inner product between two vectors,
$F=\braket{(\hat{P}_{Z_1Z_N}U)}{\hat{P}_{R}{U}}$,  
provides a simple  geometric interpretation of the perfect transport condition,
as shown  in \zfr{2}. 
The vector $\hat{P}_{Z_1Z_N}\ket{U}$ should be \textit{parallel} to  $\hat{P}_{R}\ket{U}$, which can be obtained if 
$\hat{P}_{Z_1Z_N}$ rotates $\ket{U^{\widetilde{{\cal G}}}}$  by an angle $\pi$, while  leaving 
$\ket{U^{{\cal G}}}$  unaffected. Alternatively, since $\hat P_{Z_1Z_N}$ just
describes a $\pi-$rotation of the vector $\ket{U}$ about the $Z_1Z_N$ axis,
for perfect transport the rotation-reflection operation $\hat S=\hat P_{Z_1Z_N}\hat{P}_{R}$ 
 should be a \textit{symmetry} operation for $\ket{U}$.

\begin{figure}
 \centering
\includegraphics[scale=1]{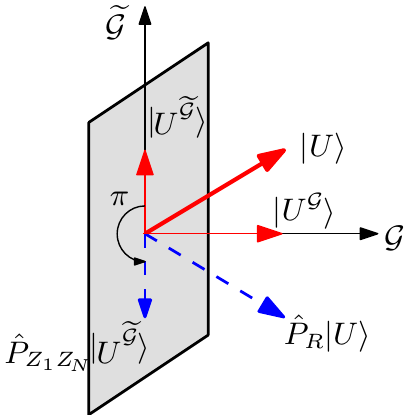}
\caption{(Color online) Geometric interpretation of the condition for maximum
$Z_1\rightarrow Z_N$ transport fidelity. The unitary $U$ is represented 
as a vector $\ket{U}$ (red thick arrow) in the HS space, with components
$\ket{U^{\cal G}}$ and $\ket{U^{\widetilde{\cal G}}}$ in the subspaces ${\cal
G}$ (represented by an axis) and $\widetilde{\cal G}$ (represented by the 
shaded grey plane). $\hat{P}_R\ket{U}$ (blue dashed arrow) is the reflection of
$\ket{U}$ about the ${\cal G}$ axis. Maximum
fidelity occurs only when the $\hat{P}_{Z_1Z_N}$ causes a $\pi$-rotation of
$\ket{U^{\widetilde{\cal G}}}$.
}
\zfl{2}
\end{figure}
From \zr{3.3} we have $\bra{U}\hat{P}_{Z_1Z_N}\ket{U}=0$ and using \zr{3.4} we can derive explicit conditions to be satisfied by the
propagator to achieve perfect transport, $F=1$
\begin{eqnarray}
 \bra{U^{\cal G}}\hat{P}_{Z_1Z_N}^{\cal G}\ket{U^{\cal G}} +
\bra{U^{\widetilde{{\cal G}}}}\hat{P}_{Z_1Z_N}^{\widetilde{\cal
G}}\ket{U^{\widetilde{{\cal G}}}}
&=&0\\
 \bra{U^{\cal G}}\hat{P}_{Z_1Z_N}^{\cal G}\ket{U^{\cal G}} -
\bra{U^{\widetilde{{\cal G}}}}\hat{P}_{Z_1Z_N}^{\widetilde{\cal
G}}\ket{U^{\widetilde{{\cal G}}}}
&=& 1
\end{eqnarray}
that simplify to
\begin{equation}
 \bra{U^{\cal G}}\hat{P}_{Z_1Z_N}^{\cal G}\ket{U^{\cal G}} =
-\bra{U^{\widetilde{{\cal G}}}}\hat{P}_{Z_1Z_N}^{\widetilde{\cal
G}}\ket{U^{\widetilde{{\cal G}}}}
=\frac{1}{2}\:.
\zl{3.5}
\end{equation}
When is this equation satisfied? By symmetry, it happens when
$\|{U}^{\cal G}\|=\|{U}^{\widetilde{{\cal G}}}\|=1/2$, and
$\ket{U^{\cal G}}$
and $\ket{U^{\widetilde{{\cal G}}}}$ are (up to a phase) eigenvectors
of
$\hat{P}_{Z_1Z_N}^{\cal G}$ and
$\hat{P}_{Z_1Z_N}^{\widetilde{\cal G}}$ with eigenvalues $\pm 1$ 
respectively:
\begin{equation}
 \hat{P}_{Z_1Z_N}^{\cal G}\ket{U^{\cal G}}=+\ket{U^{\cal G}};\quad
\hat{P}_{Z_1Z_N}^{\widetilde{\cal G}}\ket{U^{\widetilde{\cal G}}}=-\ket{U^{\widetilde{{\cal G}}}}
\zl{3.6}
\end{equation}
To enable perfect transport, $\ket{U}$ must thus have an
equal projection on the two subspaces ${\cal G}$ and $\widetilde{{\cal G}}$, as
shown geometrically in \zfr{2}. Also, intuitively from the symmetry operation $\hat S$, all components of $\ket{U}$ lying on the plane
${\cal G}$ should be rotationally symmetric with respect to $Z_1Z_N$, while
components of $\ket{U}$ lying on the plane $\widetilde{\cal G}$ should have reflection
symmetry about $Z_1Z_N$.

Note that \zr{3.6} imposes fairly weak constraints on
the transport unitaries, as opposed to the constraints for pure state transport \cite{Kostak07,Karbach05}. In
particular \zr{3.6} provides no explicit constraint on the bulk of
the network. For example, the two propagators,
\begin{equation}\begin{array}{l}
 U_1=B_{\rm bulk}(\Id  \pm Z_1Z_N)
+ B_{\rm bulk}^{'}(X_1X_N \pm Y_1Y_N),\\
 U_2=B_{\rm bulk}(\Id  \pm Z_1Z_N)
+ B_{\rm bulk}^{'}(X_1Y_N \mp Y_1X_N),
\end{array}\zl{3.7}\end{equation}
with $B_{\rm bulk}$ and $B_{\rm bulk}^{'}$ 
arbitrary equi-norm operators ($\in{\rm
span}\{\bf{B}_{\rm bulk}\}$) acting on the bulk, support perfect transport.
Other propagators can be obtained thanks
to an invariance property that we present in the next section.
More generally, in \zar{1} we explicitly provide a prescription to construct
classes of unitaries for perfect mixed state transport.

\subsection{Invariance of transport Hamiltonians}
\zsl{IIIb}
The fidelity $F$ in \zr{3.2} is invariant under a transformation $U^{'}=VU$,
where $V$ is unitary and commutes with $\hat S$, that is,
\begin{equation}
 [\hat{V},\hat{P}_{Z_1Z_N}\hat{P}_{R}] =0.
\zl{3.8}
\end{equation}

This invariance can be used to construct  Hamiltonians that
support perfect transport starting from known ones. 
Consider an  Hamiltonian $H$  that generates the transport evolution
$U=\exp(-iHt)$. 
Then the transport driven by $H$ is \textit{identical} to that generated by the
Hamiltonian $H^{'} =
V^{\dagger}HV$, where  $V$ satisfies \zr{3.8}.

Ref.~\cite{Kay11}  proved  similar symmetry requirements
for Hamiltonians that transport pure states; here, however,  we
derived these Hamiltonian properties just from the geometric conditions on $U$.
Ref.~\cite{Hincks11} treated a similar problem, defining classes of  Hamiltonians that perform the same action on a state of interest. 
A special case of this result was used  in~\cite{Cappellaro07l} to study transport in a mixed-state  spin chain  driven either by the nearest-neighbor coupling  isotropic XY Hamiltonian,  
\begin{align}
 H_{\rm XY}=\textstyle\sum_{i}\alpha_{i}T_{ii+1}^{+}\text{ with
}T_{ij}^{+}=(S_i^{+}S_j^{-} + S_i^{-}S_j^{+})
\end{align}
or  double-quantum (DQ) Hamiltonian 
\begin{eqnarray}
 H_{\rm
 DQ}=\textstyle\sum_{i}\alpha_{i}D_{ii+1}^{+}\text{ with }D_{ij}^{+}=(S_i^{+}S_j^{+} +
S_i^{-}S_j^{-}),
\zl{3.9}
\end{eqnarray}
with  $S_j^{\pm}=\frac{1}{2}(X_j \pm iY_j)$. 
The unitary operator relating the two Hamiltonians, $V =
\prod_{k^{'}}X_{k^{'}}$,
where the product $k^{'}$ extends over all even \textit{or} odd spins, does 
indeed satisfy \zr{3.8}.

\subsection{Quantum information transport via mixed state networks}
\zsl{IIIc}
The requirements for perfect transport  (\zr{3.6})
can be easily  generalized to the  transport between any
two elements of $\textbf{B}$, say from ${\cal I}$ to ${\cal F}$. 
One has simply to appropriately construct the subspaces ${\cal G}$ and
$\widetilde{\cal G}$ and
the corresponding permutation operator $\hat{P}_{{\cal IF}}$.

One could further consider under which conditions  this transport (for
example, $X_1\rightarrow X_N$) can occur \textit{simultaneously} with the $Z_1\rightarrow Z_N$ transport already
considered.
More generally, the simultaneous transfer of operators forming a basis for
$\bf{B_1}$ would enable the transport of quantum
information~\cite{Cappellaro11,Albanese04} via a mixed-state network. The unitary $U$
should now not only be symmetric under $\hat S$, but should also under a similar  operator derived for $X_1X_N$. 
The requirements on $U$ thus become more stringent 
and only a special case of the propagators constructed in \zr{A1.1} in \zar{1} is allowed,
\begin{equation}
 U=B_{\rm bulk}(\Id  \pm Z_1Z_N + X_1X_N \pm Y_1Y_N)\:.
\end{equation}
This is exactly a SWAP operation (up to  a phase) between the end-spins,  which can also lead to a transfer of arbitrary \textit{pure} states between $1$ and $N$.
Therefore, we find that perfect transport of non-commuting mixed
states between the end-spins also allows transport in
pure state networks. 
We note that quantum information could be encoded in multi-spin states~\cite{Markiewicz09,Cappellaro11} that satisfy proper symmetry conditions and thus do not impose additional conditions on the transport propagators.

\subsection{Which Hamiltonians support mixed state transport?}
\zsl{IIId}
It would be interesting to determine which Hamiltonians can generate propagators $\ket{U(t)} = \exp(-i\hat{H}t)\ket{\Id }$  for perfect transport.
Unfortunately, deriving requirements for the Hamiltonian from  the conditions on
the unitaries  is non-trivial; however, as we show below, one can still extract
useful information. 

A general Hamiltonian can  be decomposed as 
 $H = H^{{\cal G}} + H^{\widetilde{{\cal G}}}$, where $H^{{\cal G},\widetilde{{\cal G}}}$  lie
in the subspaces ${\cal G}$ and $\widetilde{{\cal G}}$, respectively. We cannot
set $H=H^{{\cal G}}$ since the Hamiltonian  {does} need to have a component  
that is non-commuting with
the target operator ($Z_N$) in order to drive the transport. 
If $H=H^{\widetilde{\cal G}}$, odd powers of $H$ are in $\widetilde{{\cal G}}$,
while even powers of
$H$ belong to ${{\cal G}}$. Then the propagator has contributions from $\ket{U^{\cal G}}$ and $\ket{U^{\widetilde{\cal G}}}$ with 
\begin{eqnarray}
 \ket{U^{{\cal G}}} &=& \ket{\Id } +
\frac{(it)^2}{2!}\hat{H}\ket{H} + \frac{(it)^4}{4!}\hat{H}^3\ket{H} +
\cdots \nonumber\\
\ket{U^{\widetilde{{\cal G}}}} &=& it\ket{H} +
\frac{(it)^3}{3!}\hat{H}^2\ket{H} + \cdots
\end{eqnarray}
We can demonstrate that in this case the Hamiltonian must satisfy two conditions  to drive perfect transport. 
First,  the ``vector'' form of the
Hamiltonian must be an eigenstate of $\hat{P}_{Z_1Z_N}$, 
$\hat{P}_{Z_1Z_N}\ket{H}=-\ket{H}$, which ensures that the second equation in (\ref{eqn:3.6})  is trivially satisfied, as $\hat{P}_{Z_1Z_N}\ket{U^{\widetilde{\cal G}}} = -
\ket{U^{\widetilde{\cal G}}}$. Second, since we have 
 \[
 \hat{P}_{Z_1Z_N}\ket{U^{{\cal G}}} = \ket{Z_1Z_N} + 
\frac{(it)^2}{2!}\hat{H}\ket{H} + \frac{(it)^4}{4!}\hat{H}^3\ket{H} +
\cdots,
\]
the first equation  in  (\ref{eqn:3.6}) 
implies that $H^{2n}\!=\!\half(\Id \!-\! Z_1Z_N)$ for any $n$. 

These conditions are for example  satisfied by  the XY-like Hamiltonian,
$H=B_{\rm bulk}T_{1N}^{+}$, where $B_{\rm bulk}$ is any operator acting on the
bulk and $T_{1N}^{\pm}=(S_1^{+}S_N^{-} \pm S_1^{-}S_N^{+})$. 
In this case, at $t=\pi/4$, all conditions in \zr{3.6} are satisfied and perfect
transport is
achieved. Indeed the XY Hamiltonian  has been widely studied for quantum
transport~\cite{Bose03,Christandl04} and it is interesting that we could derive its
transport properties solely by the symmetry conditions on the propagator.

  An Hamiltonian  $H=H^{\widetilde{\cal G}}$ with support only in $\widetilde{\cal G}$ is however a very restrictive case  
as it refers to the situation where all nodes of the network  are connected to $N$. 
 Hamiltonians with support in both subspaces are more experimentally relevant,
as they correspond to a common  physical situation, where the ends of the
network are separated in space and  direct interaction between them is zero or
too weak. 
  In the following,  we will consider this more general situation, although restricting the study to XY Hamiltonians in order to derive conditions for perfect transport.

\section{Perfect Transport in networks: Correlated Quantum walks}
\zsl{IV}
In the following, we will consider the network ${\cal N}$ to
consist of spins that are coupled by  XY-like interaction,
$\{T_{ij}^{+}\}$. We focus on this interaction since it has been shown that with appropriate \textit{engineered} coupling
strengths, $\alpha_{ij}\propto\sqrt{i(N-i)}\delta_{j,i+1}$, the XY-Hamiltonian can support perfect transport in 
linear spin chains (see e.g. \cite{Nikolopoulos04,Nikolopoulos04b,Albanese04,Benjamin03}). Thanks to the invariance
property described in \zsr{IIIb},  this analysis  applies to a much broader
class of Hamiltonians,  in particular to the  DQ Hamiltonian. 

We assume that the end spins of ${\cal N}$ are not directly coupled, thus transport needs to be mediated by the bulk of the network. 
The simplest such topology is a $\Lambda$-type configuration where the
end spins  are coupled to a single spin in the bulk. The
Hamiltonian  $\Lambda_j= (T_{1j}^{+} + T_{jN}^{+})/\sqrt{2}$,
where $j$ is a spin in the bulk, is enough to drive this transport. In this
case, $(\Lambda_j^2)^{n}\!=\!\Lambda_j^2\ \forall\,n$, and hence the
propagator is 
\begin{equation}
 U=\exp(-iHt)=\Id+[\cos(t)-1] H^2 -i \sin(t) H,
\zl{4.1}
\end{equation}
where $H^2 = 1/2[(T_{1j}^{+})^2 + (T_{jN}^{+})^2 + T_{1N}^{+}]$. Since $(\Id
- 2H^2)$ has the form of
$U_3$ in \zr{A1.1}, setting $t=\pi$ ensures $U=U_3$, yielding  perfect
transport.
This is an expected result, since this simple lambda-network  is just a 3-spin
linear chain. This result can be extended to longer chains, as long as
engineered couplings ensure that the resulting Hamiltonian is
mirror-symmetric~\cite{Karbach05,Albanese04,Wang11}.

A different situation arises when  there is more than one transport
\textit{path} possible,  
that is,  the end spins are coupled to more than one spin in the bulk with an  Hamiltonian  $H=\sum_{j\in \rm{bulk}}\alpha_j\Lambda_j$. 
For example, \zfr{3} depicts a network similar to the one considered in
\cite{Yao11,Clark05} where there are three $\Lambda$ paths
between the end-spins. 
Even if each path   \textit{individually} supports perfect transport,  evolution along different paths may not be \textit{correlated}, 
leading to destructive interference reducing the
fidelity (see \zfr{3}). 

\begin{figure}[t]
 \centering
{\includegraphics[scale=0.25]{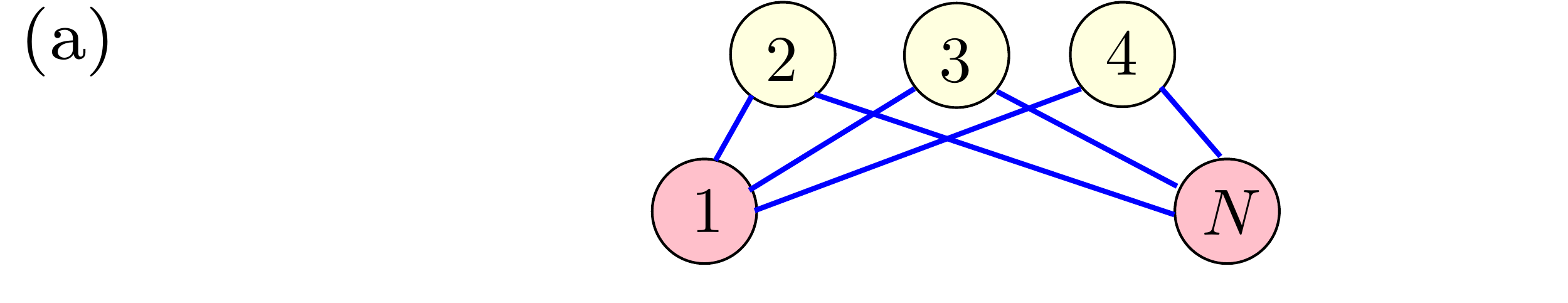}\zfl{3a}}\\
{\includegraphics[scale=0.25]{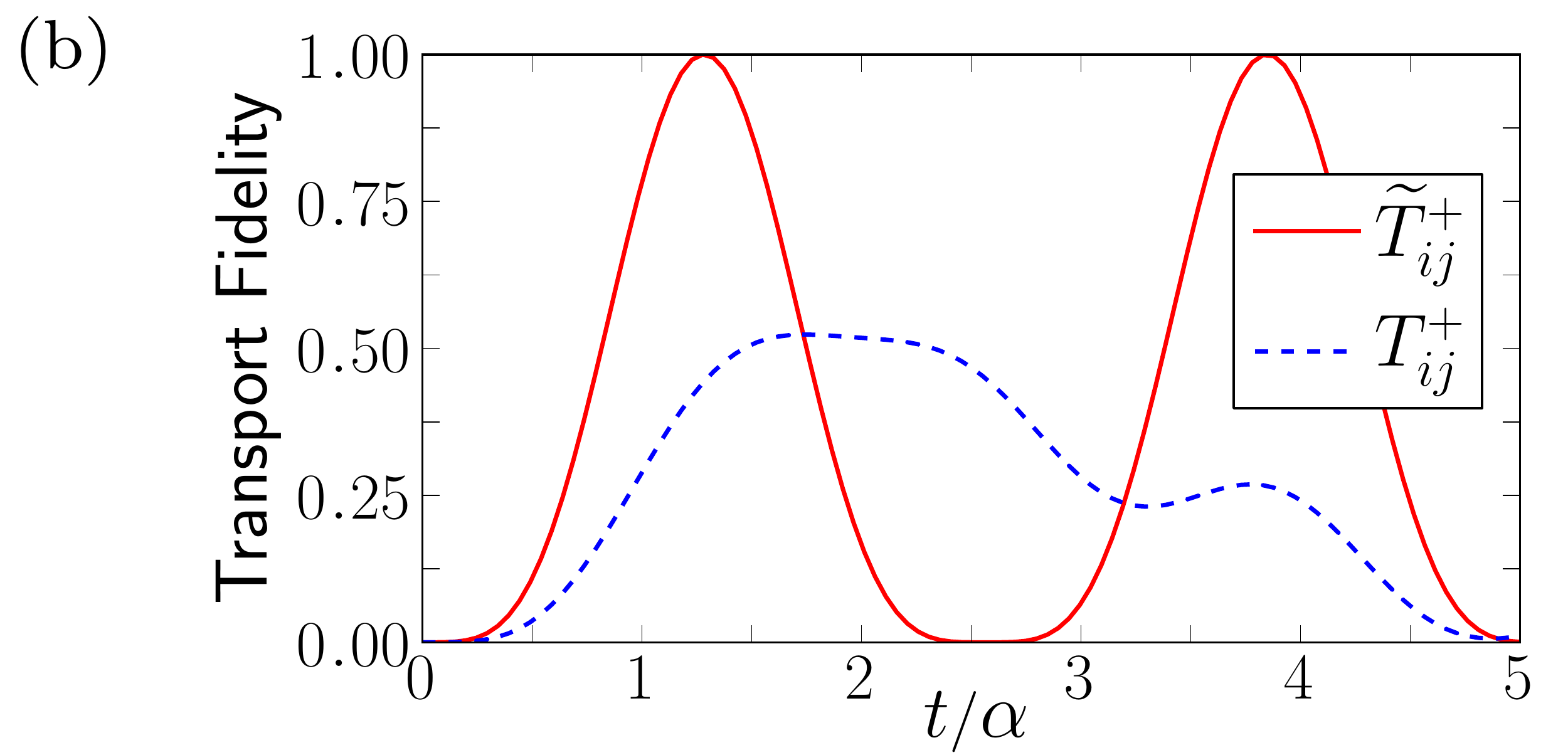}\zfl{3b}}
\caption{(Color online) (a) $\Lambda$-type network with three
$\Lambda$ paths between the bulk and end-spins and equal coupling strength $\alpha$. (b) Transport
fidelity as a function of normalized time for the $\Lambda$-network  coupled by the 
XY-Hamiltonian $T_{ij}^{+}$ (blue dashed) and the modified XY-Hamiltonian $\widetilde{T}_{ij}^{+}$
(red solid). In the latter case, correlated quantum walks lead to perfect
transport.}
\zfl{3}
\end{figure}
Perfect fidelity can be achieved only if different paths can be
\textit{collapsed} into a single ``effective'' one that  supports perfect
transport (\zfr{8}). This strategy not only allows us to determine if an
Hamiltonian can support perfect transport, but it also gives a recipe to build
allowed Hamiltonians, by combining simpler networks known to support perfect
transport into more complex ones. 

 To this end, we use the fact that linear
chains enable perfect transport with appropriate engineered couplings.
Our first step will then to give conditions under which two chains of the same
length (with end-spins in common) can be combined. To obtain these conditions,
we  describe the evolution of the spin polarization as a quantum walk over
the operators in the network \cite{Baum85,Munowitz87}. This description 
reveals the need to
{\it correlate} the parallel paths over the network, in order to achieve a
constructive refocusing of the polarization at the other end of the network. We
 then generalize the conditions  by a recursive construction to quite general networks. 

\subsection{Transport as a quantum walk over ${\cal N}$}
\zsl{IVa}
We describe the  transport evolution as a quantum walk over the network, which
progressively populates operators in the HS space. This process of progressively
populating
different parts of the HS space upon continuous time evolution under the
Hamiltonian can be considered as a quantum {\it
walk}~\cite{Baum85,Munowitz87,DiFranco07}.
We first expand the transport fidelity $F(t)$  (Eq. \ref{eqn:3.4})  in a time series, 
\begin{eqnarray}
F(t) &=&\left\langle U_0
\left|\hat{P}_{Z_1Z_N}\hat{P}_{R}\right|U_0\rangle - it\langle U_0
\left|[\hat{H},\hat{P}_{Z_1Z_N}\hat{P}_{R}] 
\right|U_0\right\rangle\nonumber\\
&+&
\frac{i^2t^2}{2!}\left\langle U_0
\left|\left[\hat{H},\left[\hat{H},\hat{P}_{Z_1Z_N}\hat{P}_{R}\right]\right] 
\right|U_0\right\rangle + \cdots
\zl{4.2}
\end{eqnarray}
with $|U_0\rangle=\ket{\Id }$. Defining the nested commutators, 
\begin{equation}
 {\cal C}_0 = \hat{P}_{Z_1Z_N}\hat{P}_{R} \ ; \ {\cal C}_n =
[\hat{H},{\cal C}_{n-1}] \:,
\zl{4.3}
\end{equation}
 \zr{4.2} takes the form 
\begin{equation}
 F(t)=\sum_{n=0}^{\infty} \frac{(it)^n}{n!}\langle{\cal C}_n\rangle,
\zl{4.4}
\end{equation}
where the expectation value is taken with respect to
$\ket{{U}_0}$.

A large part of the
Hamiltonian commutes with $\hat{P}_{Z_1Z_N}\hat{P}_{R}$ and can be neglected. 
We can isolate the non-commuting part by defining the operator $\hat{A}$ via the relationship
\begin{equation}
  [\hat{H},\hat{P}_{Z_1Z_N}\hat{P}_{R}]  =
\hat{A}\hat{P}_{Z_1Z_N}\hat{P}_{R}\:.
\zl{4.5}
\end{equation}

The operator $\hat A$ and its nested commutators ${\cal C}_n^{A}=[\hat{H},{\cal C}_{n-1}^{A}] $ (with ${\cal
C}_0^{A}=\hat{A}$) have a simple graphical construction.
The commutation relations, 
\begin{equation}
\left[{T}_{ij}^{+},{T}_{jk}^{\pm}\right] =-Z_j{T}_{ik}^{\mp};\qquad\left[{T}_{
ij } ^ { +},{T}_{kl}^{+}\right] =0\,,
\zl{4.8}
\end{equation}
(see  \zfr{4}) can be used to provide a simple prescription to graphically determine the flip-flop terms in ${\cal C}_n^{A}$. 
\begin{figure}[t]
 \centering
\includegraphics[scale=0.8]{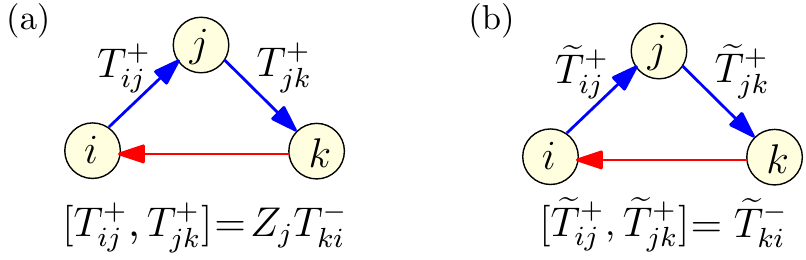}
\caption{(Color online) Graphical representation of commutators
for (a) the XY  and (b) the modified XY
Hamiltonian. The commutator between two top legs (blue) of the directed graph is
the third edge (red). In case of the XY Hamiltonian, the commutator is conditioned
on node $j$.\vspace{-12pt}
}
\zfl{4}
\end{figure}For any two edges, one in ${\cal C}_{n-1}^{A}$ and one in $H$, that share a common node,  ${\cal C}_n^{A}$ contains the edge required to complete the triangle between them. 
Thus, each higher order in the commutation expansion creates a link between
nodes in the network, progressively populating it. We will refer to the
operators ${\cal C}_n^{A}$ as quantum walk operators, since as we show below,
the nested commutators ${\cal C}_n$ in \zr{4.4} can be built exclusively out of
them.

Consider the network of \zfr{5}(a), with coupling
strengths $\alpha_{ij}=1$: $\hat{A}$ contains only the edges of ${\cal N}$ that connect to node $1$, as represented by the red lines in \zfr{5}(b). 
\begin{figure}[h]
 \centering
\includegraphics[scale=0.6]{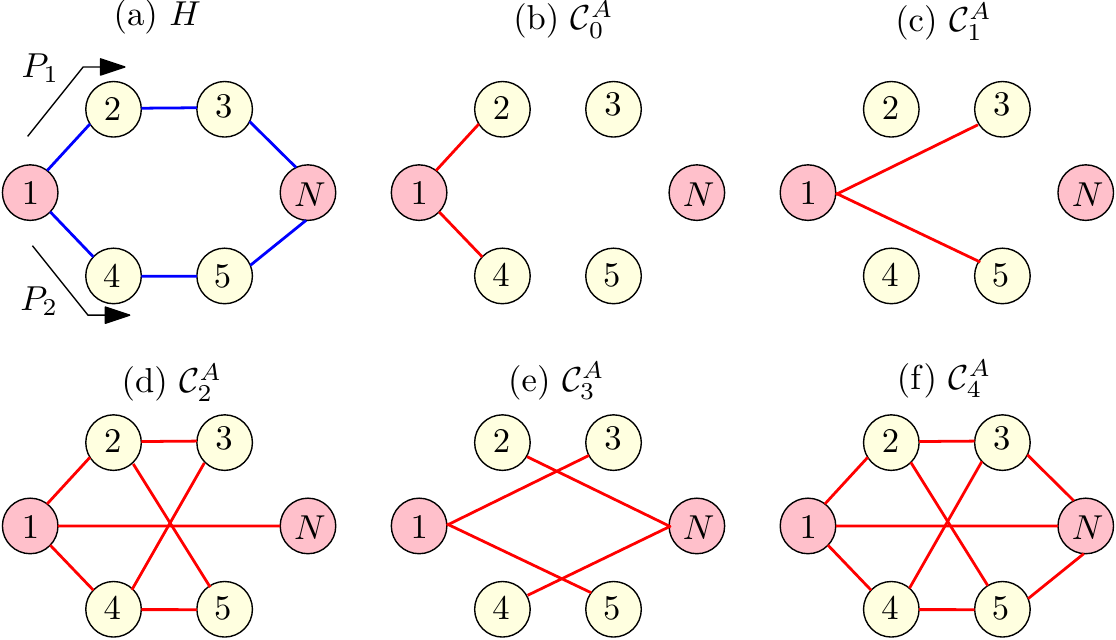}
\caption{(Color online) (a) A six spin  network with two  paths
 $P_1$ and $P_2$ between the end spins. (b)-(f) represent
graphically the successive orders of the quantum walk operators
${\cal C}_{n}^{A}$: A red line linking two
nodes indicates that there is a flip-flop term $T_{ij}^{\pm}$ between them, while  path-dependent prefactors are not depicted. 
Once the walk has
covered the entire network, successive orders
in ${\cal C}_{n}^{A}$ reproduce ${\cal C}_{3}^{A}$ and ${\cal C}_{4}^{A}$. The
explicit expressions for the commutators  are shown in \ztr{1}.
}
\zfl{5}
\end{figure}
\zfr{5}(c-d) represent the higher order commutators, with a red line linking two nodes  denoting  a term $T_{ij}^{\pm}$ between them. 
We note that the graphical construction only predicts the presence of a flip-flop term $T_{ij}^{\pm}$ linking two nodes in the commutator  ${\cal C}_n^{A}$, while the explicit forms of the commutators is generally more complex, as shown in \ztr{1}, with additional appropriate weights for arbitrary coupling strengths $\alpha_{ij}$. 
Still, as we now show, only the $T_{ij}^{\pm}$ terms are important to determine
the fidelity, and the presence of a $T_{1N}^{\pm}$ term in the graphical series
is an indication that transport can occur between the end-nodes.

The  commutators ${\cal C}_n$ can indeed be written in terms of the ${\cal C}_n^{A}$ nested commutators, 
\begin{equation}
 {\cal C}_{n}= \sum_{k=0}^{n-1}\dbinom{n-1}{k}{\cal C}_{n-1-k}^{A}{\cal
C}_{k},
\zl{4.6}
\end{equation}
yielding an expression for the fidelity  containing only products of the nested commutators ${\cal C}_{n}^{A}$:
\begin{widetext}
\begin{equation}
 {\cal C}_{n}= \sum_{k_1=0}^{n-1}\sum_{k_2=0}^{k_1-1}\cdots
\sum_{k_n=0}^{k_{n-1}-1}
\dbinom{n-1}{k_1}\dbinom{k_1-1}{k_2}\cdots \dbinom{k_{n-1}-1}{k_n}
{\cal C}_{n-k_1-1}^{A}
{\cal C}_{k_1-k_2-1}^{A}\cdots {\cal
C}_{k_{n-1}-k_n-1}^{A}\hat{P}_{Z_1Z_N}\hat{P}_{R}\:.
\zl{4.7}
\end{equation}
\end{widetext}
For a  commutator ${\cal C}_n$ to yield a non-zero contribution to the fidelity, the product of the  operators ${\cal C}_k^{A}$ should  be proportional to $Z_1Z_N$, that is, it should evaluate to even powers of $T_{1N}^{\pm}$. 
Hence very few terms appearing in \zr{4.7} actually contribute to the transfer fidelity $F$. 

\renewcommand{\arraystretch}{1.3}
\begin{table*}
\begin{tabular}{|c|c|c|}\hline\hline
{Walk Operator}   &{ XY Hamiltonian} & Modified XY Hamiltonian\\\hline
${\cal C}_{0}^{A}$   &$T_{12}^{+} + T_{14}^{+}$ &$\widetilde{T}_{12}^{+}
+ \widetilde{T}_{14}^{+}$ \\
\hline

${\cal C}_{1}^{A}$   &$Z_2T_{13}^{-} +
Z_4T_{15}^{-}$ &$\widetilde{T}_{13}^{-}
+ \widetilde{T}_{15}^{-}$ \\
\hline

${\cal C}_{2}^{A}$   
& $ T_{12}^{+}-T_{23}^{+}- Z_1Z_2T_{34}^{+}+Z_2Z_3T_{1N}^{+}$  
&$ \widetilde{T}_{12}^{+}- \widetilde{T}_{23}^{+} - \widetilde{T}_{34}^{+}+\widetilde{T}_{1N}^{+}  $\\
 & $+ T_{14}^{+} - T_{45}^{+} - Z_1Z_4T_{25}^{+} + Z_4Z_5T_{1N}^{+} $  
 &$ +\widetilde{T}_{14}^{+}- \widetilde{T}_{45}^{+}- \widetilde{T}_{25}^{+}+\widetilde{T}_{1N}^{+} $ \\
\hline

${\cal C}_{3}^{A}$   
&$(Z_4Z_5Z_N + 4Z_2)T_{13}^{-} + (Z_2Z_3Z_N + 4Z_4)T_{15}^{-} $  &
$5\widetilde{T}_{13}^{-} -4\widetilde{T}_{2N}^{-}$ \\
 & $- 2(Z_1Z_4Z_5 + Z_3)T_{2N}^{-} - 2(Z_1Z_2Z_3+ Z_5)T_{4N}^{-}$   
 &$+ 5\widetilde{T}_{15}^{-}  -4\widetilde{T}_{4N}^{-} $\\
\hline

  &$(4\Id  + Z_3Z_4Z_5Z_N)T_{12}^{+} + (4\Id  +
Z_2Z_3Z_5Z_N)T_{14}^{+} + 9(Z_2Z_3 + Z_4Z_5)T_{1N}^{+}
$  &$5\widetilde{T}_{12}^{+} + 5\widetilde{T}_{14}^{+}
+ 18\widetilde{T}_{1N}^{+}$\\

${\cal C}_{4}^{A}$   &$- (6\Id  + 3Z_1Z_4Z_5Z_N)T_{23}^{+}  - (6Z_1Z_4 + 3Z_3Z_N)T_{25}^{+}
-(6Z_1Z_2 + 3Z_5Z_N)T_{34}^{+} $   &$-9\widetilde{T}_{23}^{+} -9\widetilde{T}_{25}^{+}
-9\widetilde{T}_{34}^{+}$\\
   &$ + 2(\Id  + Z_1Z_2Z_4Z_5)T_{36}^{+}
- (6\Id  + 3Z_1Z_2Z_3Z_N)T_{45}^{+}+ 2(\Id  + Z_1Z_2Z_3Z_4)T_{56}^{+}$ 
&$+4\widetilde{T}_{36}^{+} -9\widetilde{T}_{45}^{+}
+ 4\widetilde{T}_{56}^{+}$\\
\hline\hline
\end{tabular}
\caption{Nested commutators $\cc_{n}^A$ (walk operators)  corresponding to the
graphs in \zfr{5} if the edges represent the XY Hamiltonian or the
modified XY Hamiltonian. In the first case, note the presence of path dependent
$Z_j$
prefactors, which are absent if the modified XY Hamiltonian is
used. This allows for the correlation of transport through parallel paths.}
\ztl{1}
\end{table*}

The geometric construction of  ${\cal C}_{n}^{A}$  only yields the XY operators contained in each commutator, but it does not
reflect the appearance of  prefactors $\propto Z_j$ (due to the commutator in
\zr{4.8}) that are explicitly written out in \ztr{1}. 
Thus, the geometric construction gives a necessary condition for transport, but not a sufficient one. 

The operators ${\cal C}_{n}^{A}$ describe the sum of walks over
different paths: for example in ${\cal C}_{2}^{A}$, $Z_2Z_3T_{1N}^{+}$ can be 
interpreted as the information packet reaching node $N$ through path $P_1$ in
\zfr{5}(a),
while $Z_4Z_5T_{1N}^{+}$ represents propagation via path $P_2$. 
These two terms could in principle contribute to the fidelity, as they contain $T_{1N}^{+}$. 
However, the  additional path dependent factors $\prod_k Z_k$
 lead to a loss of  fidelity. 
Transport through different paths yield different $\prod_k Z_k$ factors, resulting into a destructive ``interference'' effect. 
Note also that since different paths are weighted by different correlation factors, they cannot be canceled through some external control to recover the fidelity.
In the following section we show how a modified  Hamiltonian can {\it remove}
this
path-conditioning and thus drive perfect transport. Note that the path-dependent
factors
are as well unimportant in the case of pure states, provided the states reside
in
the same excitation manifold~\cite{Clark07,Difranco09,Cappellaro11}.

\begin{figure}[b]
 \centering
\includegraphics[scale=0.475]{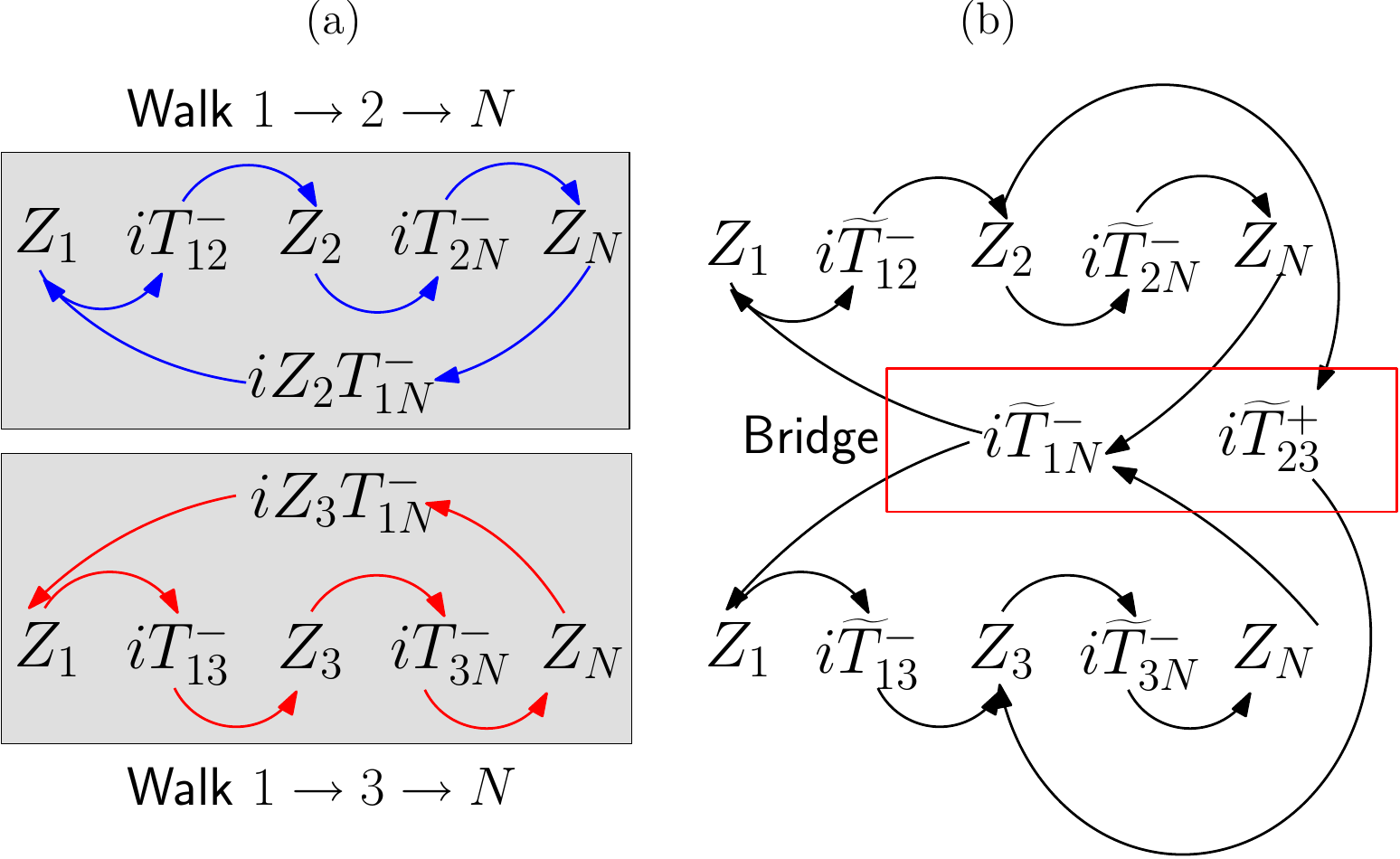}
\caption{(Color online) Operators  appearing in the quantum walk
of a  network  consisting of two $\Lambda$ paths,
$1\to 2\to N$ and $1\to 3\to N$.  In panel (a), where the transport is driven by the XY-Hamiltonian, the two paths through spins
$2$ and $3$ are different, as they traverse a different set of operators,  and are thus depicted in two separated  grey panels. In panel 
(b), where we consider  the modified XY Hamiltonian, both walks go through a common
set of operators. The previously separate walks are bridged by the operators in
the red box, making both walks indistinguishable and hence correlated.
}
\zfl{6}
\end{figure}

\subsection{Correlating quantum walks over ${\cal N}$: Modified XY-Hamiltonian}
\zsl{IVb}
\begin{figure*}
 \centering
\includegraphics[scale=0.8]{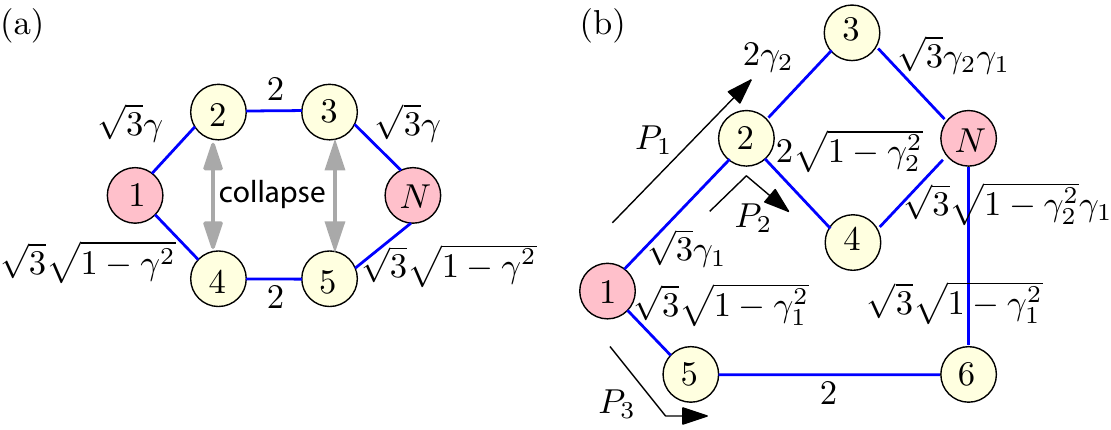}
\includegraphics[scale=0.25]{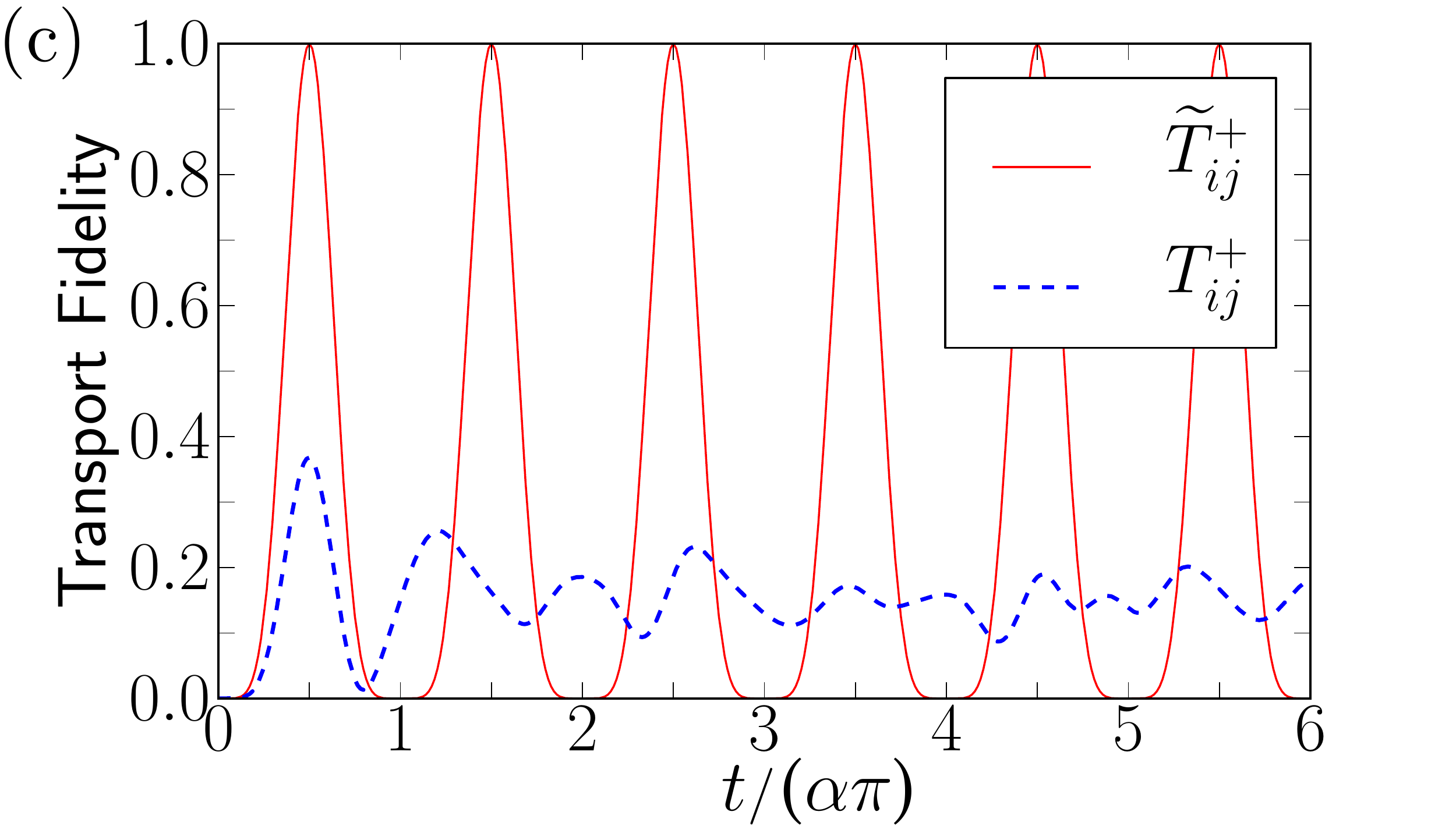}
\caption{(Color online) Engineered spin networks. (a) The
network of \zfr{5}(a), correlated by means of the modified flip-flop
Hamiltonian, can be engineered to yield perfect transport by weighting the coupling strength $\alpha$ with the coefficient shown. Here $\gamma$ is any
positive parameter, $\gamma<1$. The  two
paths $P_1$ and $P_2$ can be collapsed to form a linear chain, since nodes
$(2,4)$ and $(3,5)$ are equivalent. (b) A more complicated network consisting of
three paths $P_1-P_3$ (black arrows), with engineered strengths parametrized by
$0<\gamma_1,\gamma_2<1$. The three paths can be collapsed into an
effective 4-spin linear chain, with equivalent nodes  (2,5) and (3,4,6).
(c) Transport fidelity as a function of normalized time for the network in (b) with $\gamma_1=0.8,
\gamma_2=0.5$, in case the edges are the XY
$\{T_{ij}^{+}\}$ or modified XY
$\{\widetilde{T}_{ij}^{+}\}$ operators.  In this last case
perfect  fidelity is achieved, while for the usual XY-Hamiltonian the path-conditioned interference leads to poor transport fidelity.}
\zfl{7}
\end{figure*}

To remove the path-conditioning one should modify the Hamiltonian so that the
$Z_j$ term  in the commutator \zr{4.8} disappears. 

This can be done via a modified XY-Hamiltonian
\begin{equation}
\widetilde{T}_{ij}^{\pm}=T_{ij}^{\pm}\prod_{i<u<j}Z_u
\zl{4.10}
\end{equation}
since it satisfies this condition:
\begin{equation}
\left[\widetilde{T}_{ij}^{+},\widetilde{T}_{jk}^{\pm}\right] =-\widetilde{T}_
{ik} ^
{\mp}\:;\quad\left[\widetilde{T}_{ij}^{+},\widetilde{T}_{kl}^{+}\right] =0
\zl{4.11}
\end{equation}
These operators now depend on the number of nodes between $i$ and $j$, thus 
introducing a {\it metric} in the spin-space that distinguishes paths between the two nodes $i$ and $j$. 
Note that  when the network $\cal N$ is a simple linear chain with nearest-neighbor couplings the modified Hamiltonian $\widetilde{T}_{ij}^{+}$ is  equivalent to the bare XY Hamiltonian. 
The modification in \zr{4.10} of the  XY-Hamiltonian could also be seen as mapping the spin system into a set of non-interacting
fermions~\cite{Terhal02,Barthel09} via a Jordan-Wigner 
transformation~\cite{JW,Lieb61}, since  $C_i=\prod_{u<i}Z_uS^{+}_{i}$ are operators
that satisfy the usual fermionic anti-commutation relationships. 
When these modified operators are employed in the network ${\cal N}$ of \zfr{5}(a), 
the two paths $P_1$ and $P_2$ in \zfr{5}(a) are  {\it indistinguishable} or,
equivalently, they become perfectly correlated (see \ztr{1}). In
effect, the modified XY Hamiltonian drives the quantum walks  over
different paths through a {\it common} set of operators of \textbf{B}. This is
shown in \zfr{6} for a simple $\Lambda$-network consisting of two $\Lambda$ paths.

The graphic construction used to calculate the transport over the network in  \zfr{5} remains
unchanged, except that now the red lines between two nodes  denote modified flip-flops $\widetilde{T}_{ij}^{\pm}$
between them. Crucially there are {\it no} path dependent prefactors and  
 symmetric nodes in each path become \textit{equivalent} in each of the operators ${\cal C}_{n}^{A}$.
It is then possible to collapse different paths into a single effective
one, until a complex network ${\cal N}$ is collapsed into a linear chain. This is depicted in
\zfr{7}(a).

We can express this result more formally, by defining  \textit{collapsed} XY operators, where we denote in parenthesis
equivalent nodes in two parallel paths:
\begin{equation}
 \widetilde{T}_{i(j,k)}^{\pm} =
\frac{1}{\sqrt{\gamma_{ij}^2 + \gamma_{ik}^2}}\left(
\gamma_{ij}\widetilde{T}_{ij}^{\pm} +
\gamma_{ik}\widetilde{T}_{ik}^{\pm}\right)\:, 
\zl{4.12}
\end{equation}
where $\gamma_{ij}$ and $\gamma_{ik}$ are arbitrary parameters,
$0<\gamma_{ij},\gamma_{ik}<1$ (see also \zar{2}). Remarkably, these operators
satisfy the same
path-independent commutation relations as in \zr{4.11}
\begin{equation}
 \left[\widetilde{T}_{i(j,k)}^{+},\widetilde{T}_{(j,k)\ell}^{\pm}\right] =
-\widetilde{T}_{i\ell}^{\mp}\:,
\zl{4.13}
\end{equation}
thus showing that intermediate equivalent nodes can be neglected in higher order  commutators.
In addition the nested commutators ${\cal C}_n^A$, and the
graphical method to construct them (\zfr{4}), remain invariant when
substituting the
modified XY operator with the collapsed operators
$\widetilde{T}_{i(j,k)}^{\pm}$.

Using the collapsed operators, the network of \zfr{5}(a) can thus be reduced
to a simpler linear chain (\zfr{7}). 
Analogous arguments for path-collapsing were presented in \cite{Childs09}, and
have been applied before to  some classes of graphs \cite{Christandl04,Bernasconi08}. 
In the following we show that path-equivalence could be constructed even for more complex network
topologies, since, as we described,  path collapsing can be derived just
from the commutation relationships between the edges of the network.

\subsection{Engineered spin networks}
\zsl{IVc}
The path collapsing described in the previous section provides  a constructive way to build networks, with appropriate coupling geometries and strengths, that achieve perfect transport.
Alternatively, given a certain network geometry, the method determines all the possible coupling strength distributions that leave its transport fidelity unchanged.

For example, starting from  a linear chain, any node  can
be substituted by two equivalent nodes, thus giving rise to two equivalent
paths. Then, within the subspace of the equivalent nodes, the couplings can be
set using \zr{4.12} with arbitrary weights $\gamma$, thus giving
much flexibility in the final allowed network. 
The engineered network corresponding to \zfr{5}(a) is represented in \zfr{7}(a), where equivalent nodes from $P_1$ are weighted by $\gamma$, while those from $P_2$ are weighted by $\sqrt{1-\gamma^2}$.

A more complex example is shown in 
\zfr{7}(b), where the network is built combining the networks in \zfr{3}(a) and
\zfr{5}(a). It consists of three
paths and can be collapsed into a 4-spin linear chain. The couplings
shown lead to perfect $Z_1\rightarrow Z_N$ transport for arbitrary path weights
$\gamma_1$ and $\gamma_2$, with $0<\gamma_1,\gamma_2<1$, as shown in
\zfr{7}(c). 
The network engineering scheme can be recursively integrated to construct larger and more
complicated network topologies (see for example \zfr{8}).

Similar weighted  networks have been considered before for
bosons~\cite{Feder06}. 
The engineered couplings derived  by mapping quantum walks of $N$ spinor bosons to the walk of a single particle are however much more restricted  than what we found here via the mapping of spins to non-interacting fermions.

\begin{figure}[h]
 \centering
\includegraphics[scale=0.5]{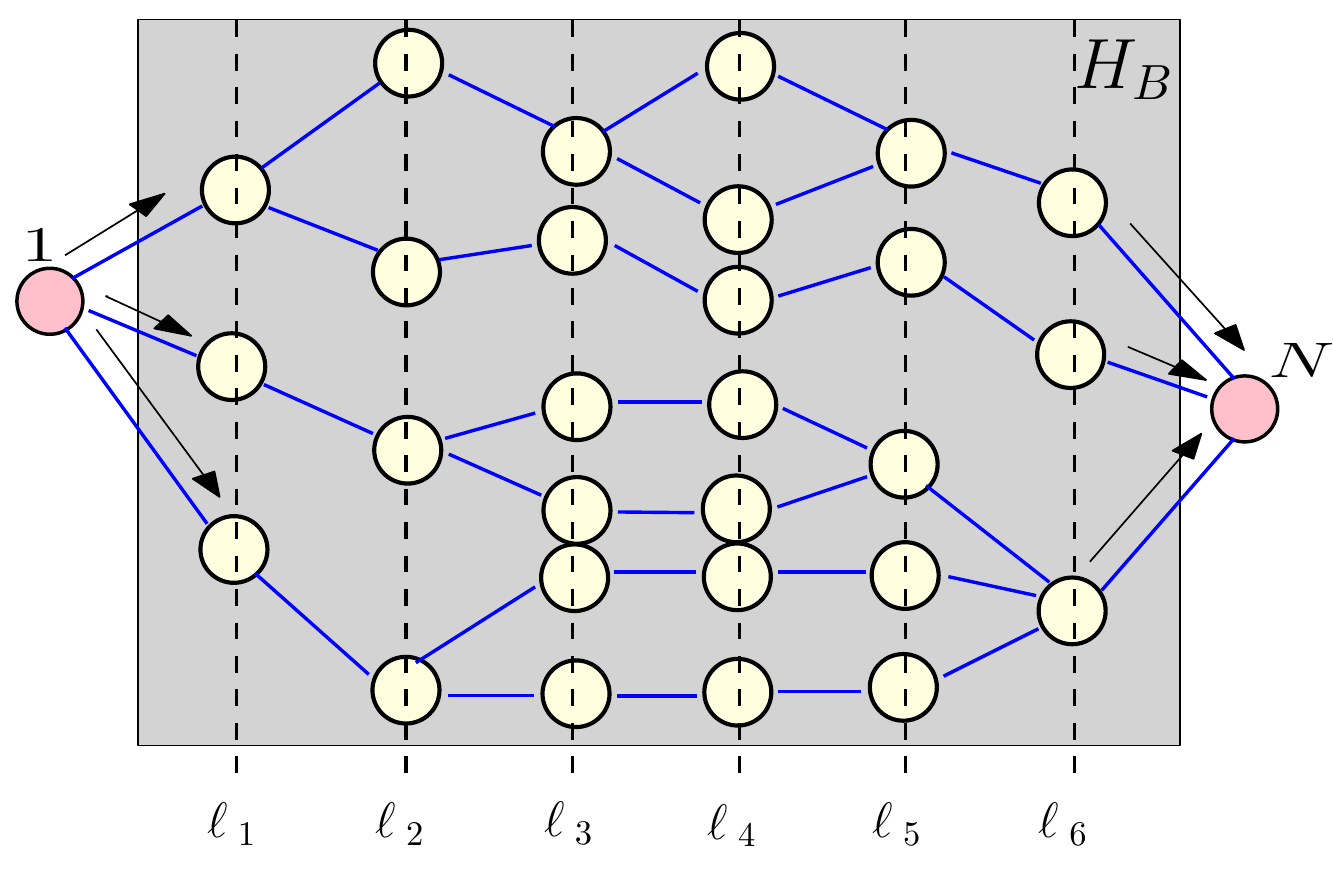}
\caption{(Color online) A  complex network topology that can be
engineered for perfect transport with the modified XY Hamiltonian.
Collapsing the equivalent nodes along the lines $\ell_2-\ell_5$
leads to
three equivalent paths that can be suitably engineered for perfect transport. 
}
\zfl{8}
\end{figure}

\section{Conclusions and Outlook}
\zsl{VI}
Experimental implementation of quantum information transport requires relaxing many of the assumptions made in ideal schemes.
 In this paper we analyzed a physical situation that is closer to experimental
settings -- information transport in mixed-state spin networks with complex
topologies. 
We first  derived general conditions on
propagators that allow  perfect  transport in these mixed-state spin networks. 
{We used the conditions on the propagators to show that there exist classes of symmetry transformations on the Hamiltonians driving the transport for which the transport fidelity is invariant. }
We also showed that the propagator conditions also imply that  transporting some mixed states requires fewer control requirements than pure state transport,  an added advantage to using mixed-state channels in quantum information architectures.

In order to study quantum transfer in complex spin networks, we described the dynamics as a continuous quantum walk over the possible paths offered by the network. This description provided  a graphical construction to predict the system evolution, which highlighted the need of correlating the transport processes occurring along different paths of the network to obtain perfect transport.  We thus introduced a modified XY-Hamiltonian, based on Jordan-Wigner fermionization, that achieves  correlation among paths by establishing a metric for the quantum walks occurring on the network. 
Conversely, the graphical construction could be as well used to study the generation from the usual XY-Hamiltonian  of states of interest in measurement-based quantum computation~\cite{Raussendorf03}.

Finally, the quantum-walk picture and the graphical construction lead us to define a constructive method to build complex networks from simpler ones, with appropriate coupling geometries and strengths, that achieve perfect transport.
We thus found that there is considerable freedom in the choice of topology and
interaction strength that still allows perfect transport in complex networks.
While the requirement of a well-defined network topology could be further
relaxed~\cite{Ajoy12x}, the precise construction proposed in this paper would
provide faster transport and the freedom in the coupling distributions could
make these networks implementable in experimental systems.

\section*{Acknowledgment}
This work was partially funded by NSF under grant DMG-1005926.

\appendix
\section{Constructing perfect transport unitaries}
\zal{1}
Here we show how the conditions specified in \zr{3.6} could be used
to construct perfect transport propagators. Our motivation for this is to
demonstrate that the conditions of \zr{3.6} are very \textit{weak},
in the sense that is possible to construct an infinite classes of unitaries
that support $Z_1\rightarrow Z_N$ transport.

Consider
the matrix forms of
$\hat{P}_{Z_1Z_N}^{\cal G}$
and $\hat{P}_{Z_1Z_N}^{\widetilde{\cal G}}$ in the two-dimensional
$\{1,N\}$ subspace of
${\cal G}$ and
$\widetilde{\cal G}$:
\begin{eqnarray}
{\cal
G}\!&\!\sim\!&\!\Big\{\{\Id ,Z_1Z_N\},\{Z_1 ,  Z_N\},
\{X_1 ,Y_1Z_N\},\{Y_1 ,X_1Z_N\}\Big\}\nonumber\\
\widetilde
{\cal G}\!&\!\sim\!&\!\Big\{\{X_1X_N,Y_1Y_N\},\{X_1Y_N,Y_1X_N\},\{  X_N,
Z_1Y_N\},\nonumber\\
&&\{  Y_N,Z_1X_N\}\Big\}
\end{eqnarray}
where the $\sim$ refers to the restriction in the $\{1,N\}$ subspace. Then, for
this order of basis, the matrix forms are block diagonal
\begin{eqnarray}
 \hat{P}_{Z_1Z_N}^{\cal G}&=&\textrm{diag}([X, X,-Y,Y]) \nonumber\\
\hat{P}_{Z_1Z_N}^{\widetilde{\cal G}}&=&\textrm{diag}([-X,
X,-Y,Y])
\end{eqnarray}
where $X$ and $Y$ are the standard Pauli matrices, whose eigenvectors with
eigenvalues $\pm 1$ are respectively
$[1,\pm1]^T$ and $[1,\pm i]^T$; this imposes a
restriction on $U$. If $B_{\rm bulk},B_{\rm bulk}^{'} \in{\rm
span}\{\bf{B}_{\rm bulk}\}$, one can explicitly list from \zr{3.6} possible
forms of $U$
for perfect transport, 
\begin{eqnarray}
 U_1&=&B_{\rm bulk}(\Id  \pm Z_1Z_N)
+ B_{\rm bulk}^{'}(X_1X_N \pm Y_1Y_N),\nonumber\\
 U_2&=&B_{\rm bulk}(\Id  \pm Z_1Z_N)
+ B_{\rm bulk}^{'}(X_1Y_N \mp Y_1X_N),\nonumber\\
U_3&=& B_{\rm bulk}(Z_1 \pm\Id Z_N)
+B_{\rm bulk}^{'}(X_1X_N \pm Y_1Y_N),\nonumber\\
U_4&=&B_{\rm bulk}(Z_1  \pm  Z_N)
+ B_{\rm bulk}^{'}(X_1Y_N \mp Y_1X_N),\nonumber\\
U_5&=& B_{\rm bulk}(X_1 \pm iY_1Z_N)
+B_{\rm bulk}^{'}(X_N \mp iZ_1Y_N),\nonumber\\
U_6&=& B_{\rm bulk}(X_1 \pm iY_1Z_N)
+B_{\rm bulk}^{'}( Y_N \pm iZ_1X_N),\nonumber\\
U_7&=&B_{\rm bulk}(Y_1 \pm iX_1Z_N)
+B_{\rm bulk}^{'}( X_N\pm iZ_1Y_N),\nonumber\\ 
U_8&=&B_{\rm bulk}(Y_1 \pm iX_1Z_N)
+B_{\rm bulk}^{'}( Y_N\mp iZ_1X_N),\nonumber\\
\zl{A1.1}
\end{eqnarray}
Note that the bulk of the network specified by $B_{\rm bulk}$ and $B_{\rm
bulk}^{'}$ can be any arbitrary operators with equal norms. In fact, the
invariance described in
\zsr{IIIb} could be used to show that the eight forms of
$U$ in \zr{A1.1} are equivalent to $U_1$ or $U_2$.

Of course, one can combine the forms
in \zr{A1.1} to form other propagators that continue to support
perfect transport. Consider for example a propagator constructed out of $U_1$ and
$U_2$ in \zr{A1.1}, with $B_{\rm bulk},B_{\rm
bulk}^{'}=\Id $
\begin{equation}
 U = \lambda_1(1\pm Z_1Z_N) + \lambda_2(X_1X_N \pm Y_1Y_N) + \lambda_3(X_1Y_N
\mp Y_1X_N)\nonumber
\end{equation}
where $\lambda_j$ are coefficients to be determined. Then, from \zr{3.3} we have
\begin{eqnarray}
\braket{U}{U}=1 &\Rightarrow& |\lambda_1|^2
+ |\lambda_2|^2 + |\lambda_3|^2 = 1\nonumber\\
 \bra{U}\hat{P}_{Z_1Z_N}\ket{U}=0 &\Rightarrow&
|\lambda_1|^2 = |\lambda_2|^2 + |\lambda_3|^2\nonumber\\
 \bra{U}\hat{P}_{Z_N}\ket{U}=0 &\Rightarrow&
{\rm Im}(\lambda_2^{\ast}\lambda_3)=0
\zl{A1.2}
\end{eqnarray}
Other conditions in \zr{3.3} are satisifed trivially.
\zr{A1.2} can be solved exactly; for example
$\lambda_j=\{1/\sqrt{2},1/2,1/2\}$ is a solution. Importantly however, if
the $B_{\rm bulk}^{'}$'s were different from each other for $U_1$ and $U_2$, the
set of equations \zr{A1.2} becomes far simpler. 

In summary, achieving $Z_1\rightarrow Z_N$ transport requires weak conditions on
the propagator driving the transport. This is as opposed to perfect pure state
transport, that requires the propagators to be isomorphic to
permutation operators \cite{Kostak07} that are mirror symmetric \cite{Karbach05}
about the end spins of the network.

\section{Properties of flip-flop  and double-quantum Hamiltonians}
\zal{2}
In this appendix, we present simple relations satisfied by the flip-flop (XY)
operators that will be used in the main paper. Note that the double-quantum
(DQ) operators in \zr{3.9} follow analogous equations. In
what follows, distinct
indices label distinct positions on the spin network unless otherwise 
specified. We start with the definition of the operators
$S$ and $E$:
\begin{equation}
 E_j^{\pm}=\frac{1}{2}(\Id  \pm Z_j)\:,\qquad S_j^{\pm}=\frac{1}{2}(X_j \pm
iY_j).
\zl{A2.1}
\end{equation}
These operators satisfy the following product rules:
\begin{equation}\begin{array}{l}
Z_jS_j^{\pm}\!=\!\pm S_j^{\pm},\quad 
(S_j^{\pm})^2\!=\!E_j^{\pm}E_j^{\mp}=0,\\
 S_j^{\pm}S_j^{\mp}\!=\!(E_j^{\pm})^2=E_j^{\pm}.
\end{array}\end{equation}
We define the {flip-flop} operators $T_{ij}^{\pm}$ and $L_{ij}^{\pm}$:
\begin{equation}
 T_{ij}^{\pm}=(S_i^{+}S_j^{-} \pm
S_i^{-}S_j^{+});\quad L_{ij}^{\pm}=(E_i^{+}E_j^{-} \pm
E_i^{-}E_j^{+})
\zl{A2.2}
\end{equation}
From the definition in \zr{A2.2} it follows that
\[
 T_{ij}^{\pm}=\pm T_{ji}^{\pm}\:;\:Z_jT_{ij}^{+}=T_{ij}^{-}.
\]
We have then the following product relations: 
\begin{equation}\begin{array}{l}
 \left(T_{ij}^{\pm}\right)^2=\pm L_{ij}^{+}\,,\quad T_{ij}^{\pm}L_{ij}^{+}
=L_{ij}^{\pm}T_{ij}^{+}=T_{ij}^{\pm},\\
\left(L_{ij}^{\pm}\right)^2=L_{ij}^{+}\,,\quad
T_{ij}^{\pm}T_{jk}^{+}=\frac{1}{2}\left(T_{ik}^{\pm}-Z_jT_{ik}^{\mp}
\right)
\end{array}
\zl{A2.3}\end{equation}
and the  commutation relations: 
\begin{equation}\begin{array}{ll}
\left[T_{ij}^{+},T_{jk}^{+}\right] =-Z_jT_{ik}^{-}\,,\ &\left[T_{ij}^{+},
Z_jT_{ik}^{+}\right] =T_{kj}^{-},\\
\left[T_{ij}^{+},Z_i\right] = -2T_{ij}^{-}\,,&
\left[T_{ij}^{-},Z_i\right] = -2T_{ij}^{+}.
\end{array}
\zl{A2.4}\end{equation}

We define the \textit{modified} flip-flop operators
$\widetilde{T}_{ij}^{\pm}$,
\begin{equation}
\widetilde{T}_{ij}^{\pm}=T_{ij}^{\pm}\prod_{i<u<j}Z_u,
\zl{A2.5}
\end{equation}
obtained by multiplying the flip-flop operator in \zr{A2.2} by a factor of $Z_u$ for all nodes between $i$ and $j$. 
The modified flip-flop operators follow especially simple commutation rules 
\begin{equation}
\left[\widetilde{T}_{ij}^{+},\widetilde{T}_{jk}^{\pm}\right] =
-\widetilde{T} _{ik}
^{\mp}\:;\:\left[\widetilde{T}_{ij}^{+},\widetilde{T}_{k\ell}^{\pm}\right] =0
\zl{A2.7}
\end{equation}
Note that crucially, these commutators only depend on the initial and
final nodes ($i$ and $k$), and are independent of intermediate nodes. In a
physical analogy, the modified operators $\widetilde{T}_{ij}^{+}$ behave as
if they were \textit{path independent}.
Thus, when considering two (or more) paths, we could  omit any intermediate node, since it would not enter in the ensuing commutators. We then denote $\textit{equivalent}$ nodes in parenthesis --for example, $(j,k)$
means nodes $j$ and $k$ are equivalent-- and  define the \textit{collapsed}
operators:
\begin{equation}
 \widetilde{T}_{i(j,k)}^{\pm} =
\frac{1}{\sqrt{\gamma_{ij}^2 + \gamma_{ik}^2}}\left(
\gamma_{ij}\widetilde{T}_{ij}^{\pm} +
\gamma_{ik}\widetilde{T}_{ik}^{\pm}\right),
\zl{A2.8}
\end{equation}
where $\gamma_{ij},\ \gamma_{ik}$ are arbitrary parameters,
$0\!<\!\gamma_{ij},\!\gamma_{ik}\!<~\!1$. The collapsed operators satisfy 
commutation relations similar to \zr{A2.7}:
\begin{equation}\begin{array}{l}
\left[\widetilde{T}_{i(j,k)}^{+},\widetilde{T}_{(j,k)\ell}^{\pm}\right]\!=\!
-\widetilde{T}_{i\ell}^{\mp}\,,\ \  
\left[\widetilde{T}_{(j,k)i}^{+},\widetilde{T}_{i\ell}^{\pm}\right]\!=\!
-\widetilde{T}_{(j,k)\ell}^{\mp},\\
\displaystyle\left[\widetilde{T}_{(j,k)i}^{+},\widetilde{T}_{i(m,n)}^{\pm}\right] =
-\widetilde{T}_{(j,k)(m,n)}^{\mp}
\end{array}
\zl{A2.9}\end{equation}
The collapsed operators in \zr{A2.8} can be generalized. If
$\mathbf{I}=(a_1,a_2,\cdots,a_m)$ and ${\bf J}=(b_1,b_2,\cdots,b_n)$ denote
two sets of equivalent nodes, we have the collapsed operator
\begin{equation}
 \widetilde{T}_{{\bf IJ}}^{\pm} =
\frac{1}{\sqrt{\sum_{i=1}^{m}\sum_{j=1}^{n}\gamma_{a_ib_j}^2}}
\sum_{i=1}^{m}\sum_{j=1}^{n}\gamma_{a_ib_j}\widetilde{T}_{a_ib_j}^{\pm}\:, 
\zl{A2.10}
\end{equation}
which satisfies the commutation relationships:
\begin{equation}
\left[\widetilde{T}_{\mathbf{IJ}}^{+},\widetilde{T}_{\mathbf{JK}}^{\pm}\right] =
-\widetilde{T} _{\mathbf{IK}}
^{\mp}\:;\:\left[\widetilde{T}_{\mathbf{IJ}}^{+},\widetilde{T}_{\mathbf{KL}}^{
\pm }\right ]=0.
\zl{A2.11}
\end{equation}

%

\end{document}